\documentclass[journal=jpcbfk,manuscript=article]{achemso}

\usepackage[version=3]{mhchem} 
\usepackage[T1]{fontenc}       

\usepackage{amsmath}
\usepackage{graphicx}
\usepackage{epstopdf}
\usepackage{bpchem}
\usepackage{color}
\usepackage[textwidth=5cm,textsize=footnotesize]{todonotes}
\setkeys{acs}{articletitle = true}

\usepackage{url}
\definecolor{tred}{RGB}{165,0,33}
\definecolor{tblue}{RGB}{0,128,255}
\definecolor{tgrey}{RGB}{128,128,128}

\author{Tatsiana Burankova}
\affiliation[Tartu University]{Institute of Physics, University of Tartu, Tartu, Estonia}
\alsoaffiliation[Paul Scherrer Institute]{Laboratory for Neutron Scattering and Imaging, Paul Scherrer Institute, Villigen PSI, Switzerland}
\email{tatsiana.burankova@psi.ch}

\author{Thomas Hau{\ss}}
\affiliation[HZB]{Helmholtz-Zentrum Berlin, Germany}

\author{Jacques Ollivier}
\affiliation[ILL]{Institut Laue--Langevin, Grenoble, France}

\author{Ruep E. Lechner}
\affiliation[ESS]{European Spallation Source, Lund, Sweden}

\author{Norbert A. Dencher}
\affiliation[TUD]{Technische Universit\"at Darmstadt, Germany}
\alsoaffiliation[RMM]{Research Center for Molecular Mechanisms of Aging and Age-related Diseases, Moscow Institute of Physics and Technology, Dolgoprudny, Russia}

\author{J\"org Pieper}
\affiliation[Tartu University]{Institute of Physics, University of Tartu, Tartu, Estonia}
\email{jorg.pieper@ut.ee}

\title{Data Analysis for Time-Resolved QENS Experiments in Biophysics}

\abbreviations{QENS, PM, BR}

\begin{document}
\begin{abstract}

To study the correlation between internal protein dynamics and protein functionality of bacteriorhodopsin, we have performed time-resolved quasielastic neutron scattering (QENS) experiments combined with in situ laser activation of protein function. A set of purple membrane samples with different lamellar lattice constants and, hence, different hydration levels has been examined at different time delays between the laser and neutron pulses. While the water content affected the lineshapes of QENS spectra, the laser-induced response remained constant within the accuracy of the method. 


\end{abstract}

\section{Introduction}

Protein dynamics span over many decades in time and are a subject of numerous studies by various techniques, as the relationship between protein internal motions and function remains in the focus of major interest~\cite{Lewandowski2015, HenzlerWildman2007, Karplus2002, Saito2014}. Although the importance of dynamical effects has been questioned~\cite{Kamerlin2010}, the search for functional-dynamical correlation has resulted in examples, which demonstrated that specific molecular motions and biological activity were concurrently suppressed or ceased because of  temperature or hydration decrease~\cite{Pieper2007, Doster2005, Ostermann2000, Rasmussen1992}. 

A seven-helix membrane protein bacteriorhodopsin (BR) is the best-studied light-driven proton pump~\cite{Ernst2014, Haupts1999} and, therefore, a convenient model system to investigate the interplay of functional and thermal motions of the protein and its environment. Moreover, the research in this field is stimulated by potential usage of BR in optoelectronics and in sensor applications~\cite{Knoblauch2014}. For example, it was observed that the conformational rearrangements in BR and its flexibility enhance the efficiency of photoconductance~\cite{Mukhopadhyay2014}.

Absorption of a light quantum by the retinal triggers a cascade of well-defined protein-chromophore conformations or intermediates with lifetimes in various time domains~\cite{Ernst2014, Lanyi2006}. The accurately described proton-exchange steps between the key residues after the photoisomerization of the retinal are accompanied by structural rearrangements in BR, which have been examined by a variety of neutron, X-ray and electron diffraction experiments and high-speed atomic force microscopy~\cite{Dencher1989, Hirai2009, Hirai2009a, Andersson2009, Wickstrand2015, Shibata2010}. The most pronounced changes in helical tilts are observed during the M intermediate, which is widely accepted to be the key step for vectorial proton transport~\cite{Ernst2014}. 

Quasielastic neutron scattering (QENS)~\cite{Bee1988, Fitter2006} has also delivered results on function-dynamic correlation in BR~\cite{Gabel2002, Lechner2011}. Being extremely sensitive to ubiquitous in biological system hydrogen atoms, QENS probes stochastic motions on the picosecond-nanosecond time scale and has an advantage of obtaining both spatial and time characteristics of the processes simultaneously. For instance, it was determined that the suppression of diffusive protein motions in BR and cessation of the proton transport occur together below 230 K and at hydration levels lower than 70\% relative humidity (r.h.).~\cite{Fitter1999}. Except varying external parameters, it became possible to directly observe modulated protein picosecond dynamics during the BR photocycle by applying a novel type of the laser-neutron pump-probe experiment~\cite{Pieper2008, Pieper2009}. Thus, a temporary increase in protein flexibility was demonstrated to occur in the M intermediate. However, due to the technical limitations the analysis had to be carried out for averaged over all detectors spectra, or to put it differently, the spatial information was traded for better statistics. This fact determined the aim of the following study, where the spectra were analyzed in the whole experimentally available ($E,Q$)-domain. Moreover, the laser-induced response of BR was studied in samples with different hydration level.

\section{Experimental Details}

\section{Sample}

%
%
%
Purple membrane stacks were prepared as described previously~\cite{Lechner2011}. D$_2$O was used to suppress the incoherent scattering from the solvent and to focus on the internal dynamics of the protein-lipid complex. In PM non-exchangeable hydrogen atoms of BR make up 75\% of the total number of the protons~\cite{Fitter1999}, so that the predominant part of the measured incoherent signal is due to the BR contribution. The in-plane lattice constant $d$ were determined on the V1-membrane diffractometer of the research reactor of the Helmholtz Zentrum f\"ur Materialien und Energie (HZB, Berlin, Germany)~\cite{Hauss2016}. The data were collected at a wavelength of 4.53 \AA{}. Different lamellar lattice constants (from 62.4 to 96 \AA{}) correspond to different amount of heavy water in the sample due to various conditions for sample equilibration~\cite{Lechner1998}.

%
%
%

\section{QENS Experiment}

The principle of a time-resolved QENS experiment has been presented in the previous publications~\cite{Pieper2008, Pieper2009}. The current measurements were performed on IN5 at Institute Laue-Langevin (ILL), Grenoble, France. QENS spectra of PM samples were recorded with a time delay of $t_{i+1}-t_i = 7060\;\mu s$ between individual slices (56 slices in total), the first spectrum corresponded to the time delay of 1000 $\mu$s between the laser and neutron pulses.

\begin{figure}[h]
\centering
\includegraphics[width=\textwidth, clip, trim=0cm 0cm 0.0cm 0.0cm]{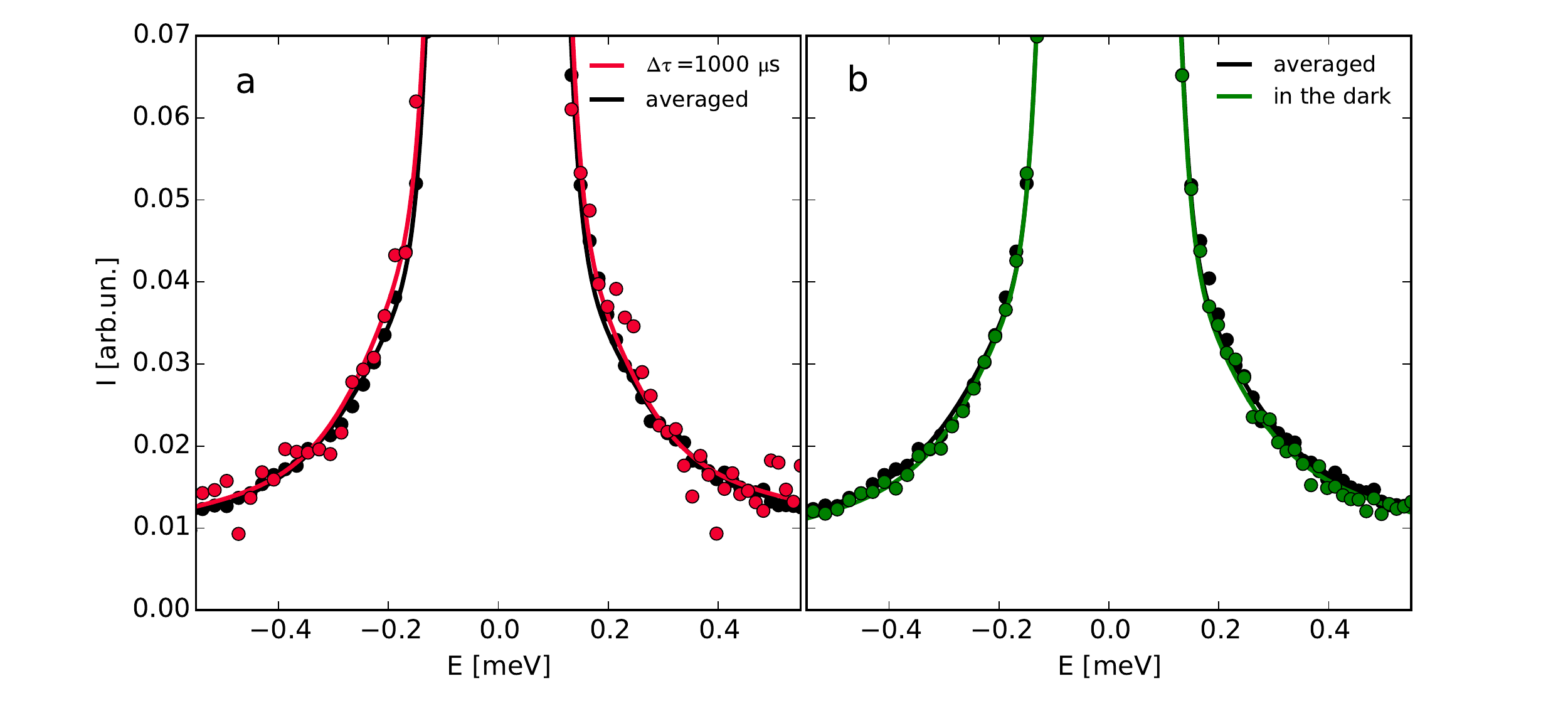}
\caption{Comparison of the QENS spectra at $Q$ = 2.0 \AA$^{-1}$ averaged over time (a, b, black line) with the spectrum recorded at the time delay of 1000~$\mu$s (a, red line) and with the spectrum measured in the dark conditions (b, green line).  }
\label{QENS_spectra}
\end{figure}

The sample temperature was maintained constant and equal to 297.0$\pm$0.2~K. Within the margin of error, the temperature of the illuminated samples did not differ from the temperature of the samples measured in the dark conditions. This can also be inspected by comparing their QENS-spectra (Figure~\ref{QENS_spectra}). Thus, the pulse frequency of 2.5~Hz was sufficient for the light energy delivered to the system via the excitation of the retinal to dissipate in the system in the course of the time period between two laser flashes. According to our estimation based on comparison of the spectra of the non-illuminated and illuminated samples, the thermal equilibration time does not exceed 40-50 ms. A similar value can be estimated by solving the heat equation numerically. The dissipation of heat in the sample holder occurs largely through the sapphire windows and the helium atmosphere. 
However, the limiting factor in this case is the thermal diffusivity of the sample, which is comparable to the corresponding value of water. It leads to the temperature decay with the half time of 35~ms for a typical sample thickness of 15--20~$\mu$m. This value is much larger than the rise time of the M intermediate, so that we could observe alterations related to the photocycle and maintain the same sample temperature over 6 hour for a single measurement.   

The wavelength of incident neutrons equaled 5.0 \AA{}, the corresponding settings of the instrument providing an elastic energy resolution function of 90~$\mu$eV and an accessible $Q$-range of 0.6 -- 2.2 \AA{}$^{-1}$. The efficiency of the detectors was calibrated by measuring a vanadium standard. The vanadium spectra were also used as the resolution function $R(Q,E)$ of the instrument. Empty cell runs were performed for further background subtraction.


%

\section{Data Analysis}

The standard data reduction of the IN5 data was performed using the LAMP~\cite{lamp} software package. The raw data were corrected for empty cell contribution, sample geometry-dependent self-attenuation and detector efficiency, converted to energy scale and finally binned into several $Q$-groups with $\Delta Q = 0.2$ \AA{}$^{-1}$ to ensure adequate data statistics.

The simultaneous fitting of all slices was performed for the whole accessible ($E$, $Q$)-space in a
program module~\cite{BurankovaPhD} based on the MPfit procedure~\cite{Markwardt2009}. This approach has a statistical advantage over the analysis of individual $Q$-groups, which was applied only as a preliminary step to determine possible theoretical scattering functions describing the studied system. The program module~\cite{BurankovaPhD} was updated to perform automatic fits of all 56 spectra measured with different time delays ($t_1=1000\;\mu s$, $t_{i+1}-t_i = 7060\;\mu s$) after the laser excitation. 

\begin{figure}[h]
\centering
\includegraphics[width=0.9\textwidth, clip, trim=0cm 0cm 0.0cm 0.0cm]{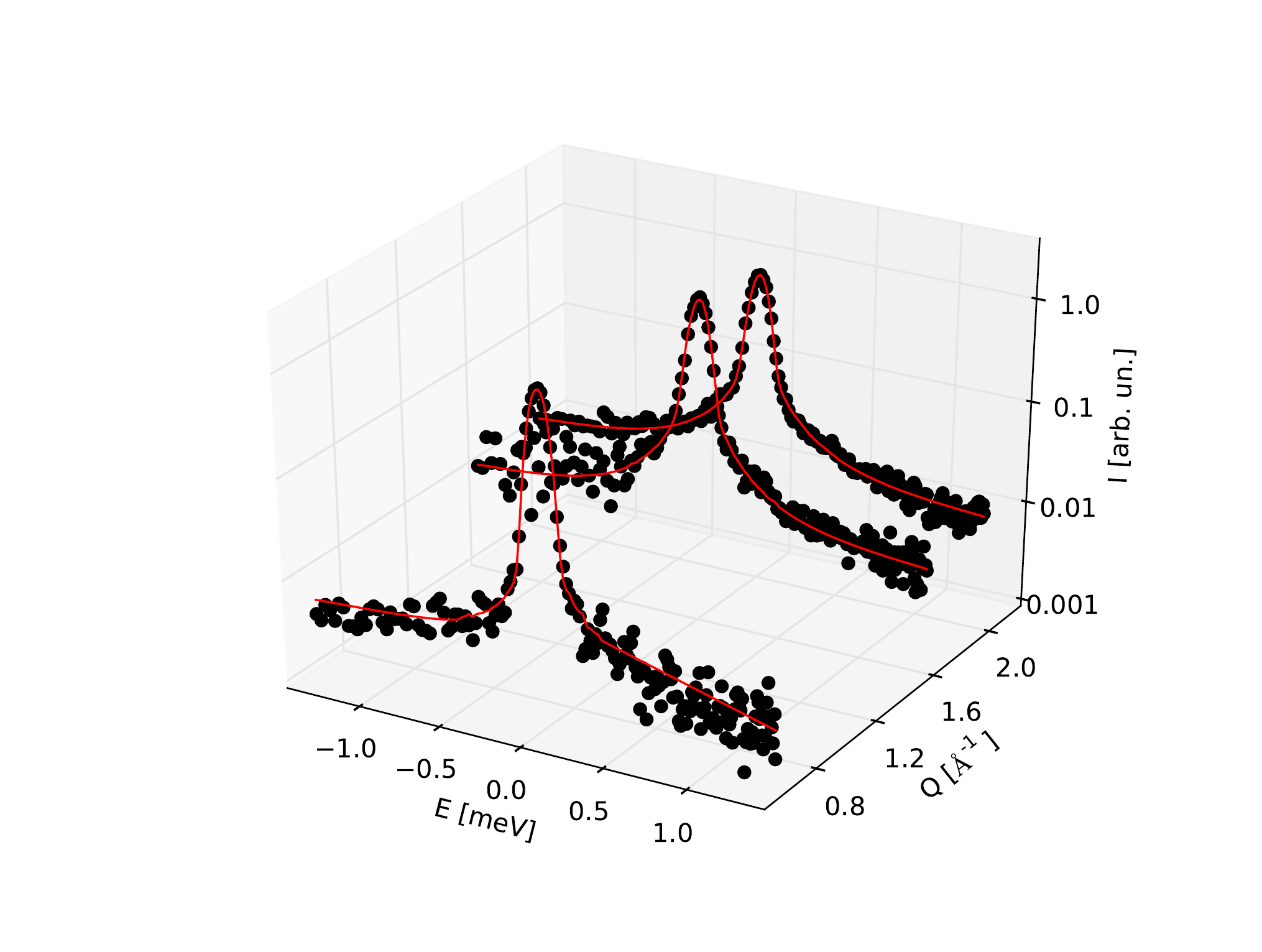}
\caption{QENS spectra of a PM sample with the fit curves obtained as a result of the 2D-fitting routing according to eq~\ref{RotSQE}, \ref{SQE2} for selected $Q$-groups}
\label{SimFit3groups}
\end{figure}

The origin of the QENS broadening in proteins is structural fluctuations occurring irrespective of the nature of the solvent and solvent-plasticized motions of side-chains~\cite{Doster2005}, which lead to a broad and continuous distribution of linewidths. Moreover, in the case of a membrane protein one has to deal with the coupled protein/lipid and solvent dynamics~\cite{Tobias2009}. Nevertheless, in terms of the phenomenological approach, a satisfactory fit of the PM data usually requires only a delta function and two quasielastic components in the quasielastic dynamical range~\cite{Pieper2008, Pieper2009}: 

\begin{equation}
\label{phenSQE}
S_\text{phen}(Q, E) = I_0(Q)\cdot\left\{A_0(Q)\delta(E)+A_1(Q)\mathcal{L}(E,\Gamma_1)+A_2(Q)\mathcal{L}(E,\Gamma_2)\right\} + bk(Q)
\end{equation}

\noindent where $\mathcal{L}(E,\Gamma_i)$ are the Lorentzian components with the half width at half maximum (HWHM) $\Gamma_i$, $I_0(Q)\sim \exp(-\langle u^2\rangle Q^2)$ is the intensity factor containing the Debye-Waller factor. The terms $A_0(Q)$ and $A_i(Q)$ stand for the elastic and quasielastic incoherent structure factors (EISF, QISF), respectively. Motions occurring on the time scale faster than several picoseconds and giving rise to a quasielastic contribution broader than the given energy transfer window are taken into account by introducing the flat background term, $bk$. Used just as the first step in our analysis, this approach, nevertheless, allowed us to notice that $A_0(Q)$ values tend to a constant value of 0.55-0.60 for $Q>1.8$ \AA{}$^{-1}$. EISF, or $A_0(Q)$ provides insight into the geometry of the spatial confinement, in which localized processes occur~\cite{Bee1988, Fitter2006}, and such a behavior serves as an indication of a certain fraction of immobile particles $p_\text{mob}$, which do not contribute to the overall QENS broadening characterized by the two Lorentzian contributions. The $Q$-dependencies of the linewidths $\Gamma_1$ and $\Gamma_2$ are not typical for unrestricted diffusion laws (self-diffusion, jump-diffusion, etc~\cite{Bee1988, Hempelmann2000}) as well. Therefore, the most common form of the scattering function can be presented as follows: 

\begin{equation}
\label{SQE1}
S(Q,E)=I_0(Q)\cdot\left\{\left(1-p_\text{mob}\right)\delta(E)+p_\text{mob}S_\text{conf}(Q,E)\right\}+bk(Q)
\end{equation}

\noindent where $S_\text{conf}(Q,E)$ is the dynamic structure factor related to the confined dynamics of flexible side chains in PM. In order to utilize all the advantages of the simultaneous analysis in the two-dimensional space $(Q,E)$, the total number of free parameters should be reduced and the $Q$-dependent terms in eq~\ref{SQE1} should have explicit analytical expressions. There are several general models applicable for describing localized stochastic motions in biological macromolecules: rotational diffusion, diffusion in a sphere~\cite{Bee1988, Hempelmann2000} or Gaussian model~\cite{Volino2006}. The background contribution can be approximated by a quadratic dependence ($bk(Q) = b + kQ^2$) irrespective of the model chosen for $S_\text{conf}(Q,E)$. After the $\chi^2$-values of the fits with the different model functions had been compared, we decided in favour of the continuous rotation diffusion model for our further analysis:

\begin{equation}
\begin{split}
S_\text{R}(Q,E)=&j_0^2(QR)\delta(E)+\\
+&\sum_{k=1}^{\infty}(2k+1)j_k^2(QR)\frac{1}{\pi}\frac{k(k+1)\tfrac{\hbar}{6\tau_\text{R}}}{\left(k(k+1)\tfrac{\hbar}{6\tau_\text{R}}\right)^2+E^2}
\end{split}
\label{RotSQE}
\end{equation}

\begin{equation}
S(Q,E)=I_0(Q)\cdot\left\{(1-p_\text{mob})\delta(E)+p_\text{mob}S_\text{R}(Q,E)\right\} + bk(Q)
\label{SQE2}
\end{equation}

\noindent where $j_\text{k}(x)$ is the k-th order spherical Bessel function. $R$ and $\tau_\text{R}$ denote the radius of the sphere and the relaxation time for rotation, respectively. These parameters have the meaning of the averaged values, because it is an evident simplification to characterize all the variety of internal motions in the system with a few independent variables. The first five terms of the infinite series in eq~\ref{RotSQE} were used for calculations, because the higher order terms converge quickly to zero in the explored $Q$-range. The typical results of the 2D-fitting procedure are displayed in Figure~\ref{SimFit3groups}.

At this point it is necessary to mention that any laser-induced effects are less pronounced than the differences in the fit parameters caused by variations in hydration level of different samples, which were at our disposal. Therefore, we decided not to sum up all the QENS spectra, but instead to analyse them separately and to compare only the deviations of the key parameters after laser excitation. Because the scatter of the fit parameters from slice to slice turned out to be significantly larger than the standard 1-sigma value derived from the covariance matrix during the minimisation procedure of an individual slice, all the errors presented here were calculated as $2\sigma$, where $\sigma$ is the standard deviation of corresponding parameters over slices 10--56, or for time delays larger than 71 ms.

\subsection{Results and Discussion}

\begin{figure}[p]
\centering
\includegraphics[width=0.7\textwidth, clip, trim=0cm 0cm 0.0cm 0.0cm]{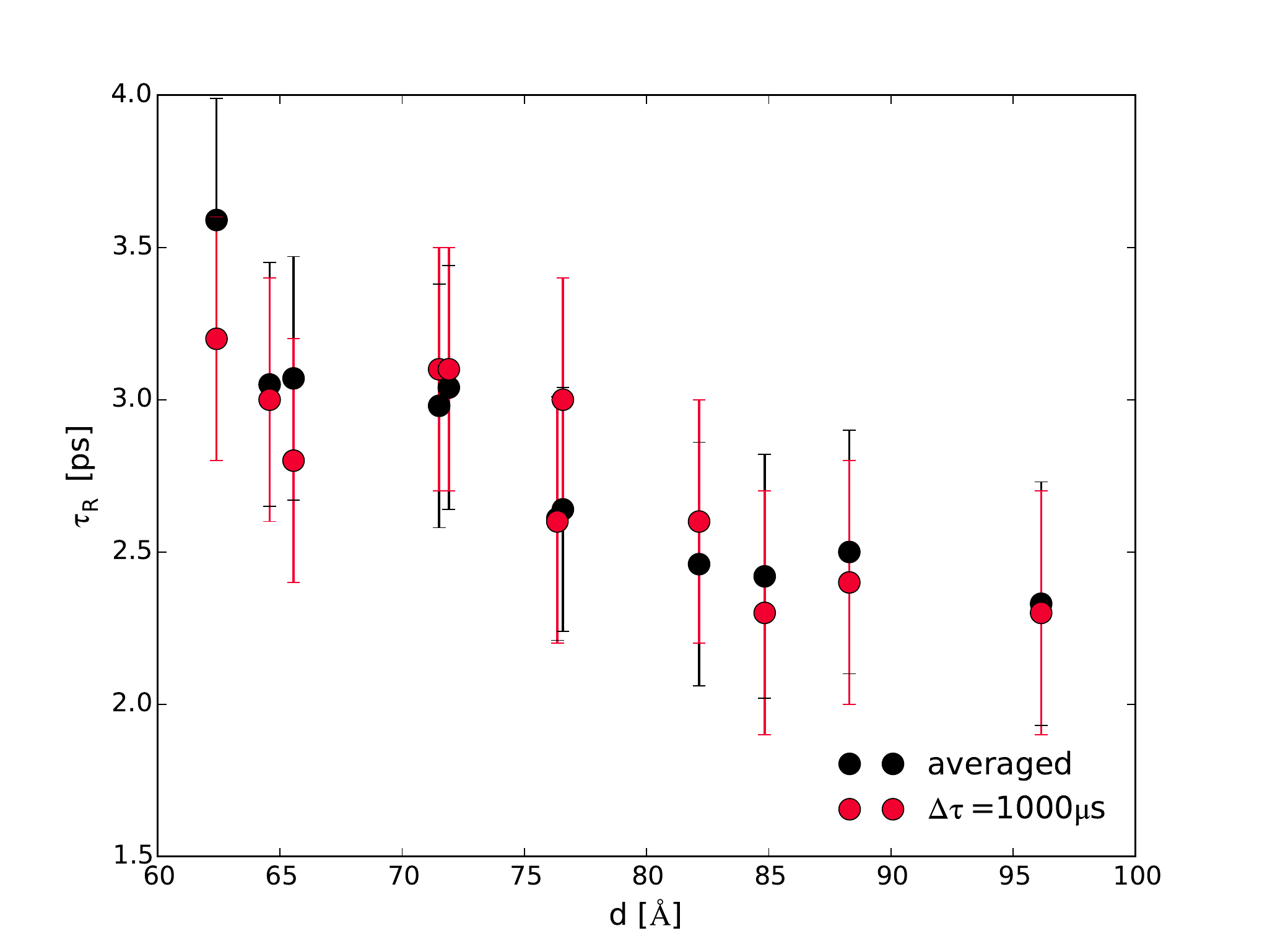}
\caption{Relaxation time for rotation as a function of the lamellar lattice constant. The black symbols present the values averaged for time delays larger than 71~ms, whereas the red symbols are the relaxation times obtained at 1000~$\mu$s after the excitation pulse.}
\label{tau_d}

\includegraphics[width=0.7\textwidth, clip, trim=0cm 0cm 0.0cm 0.0cm]{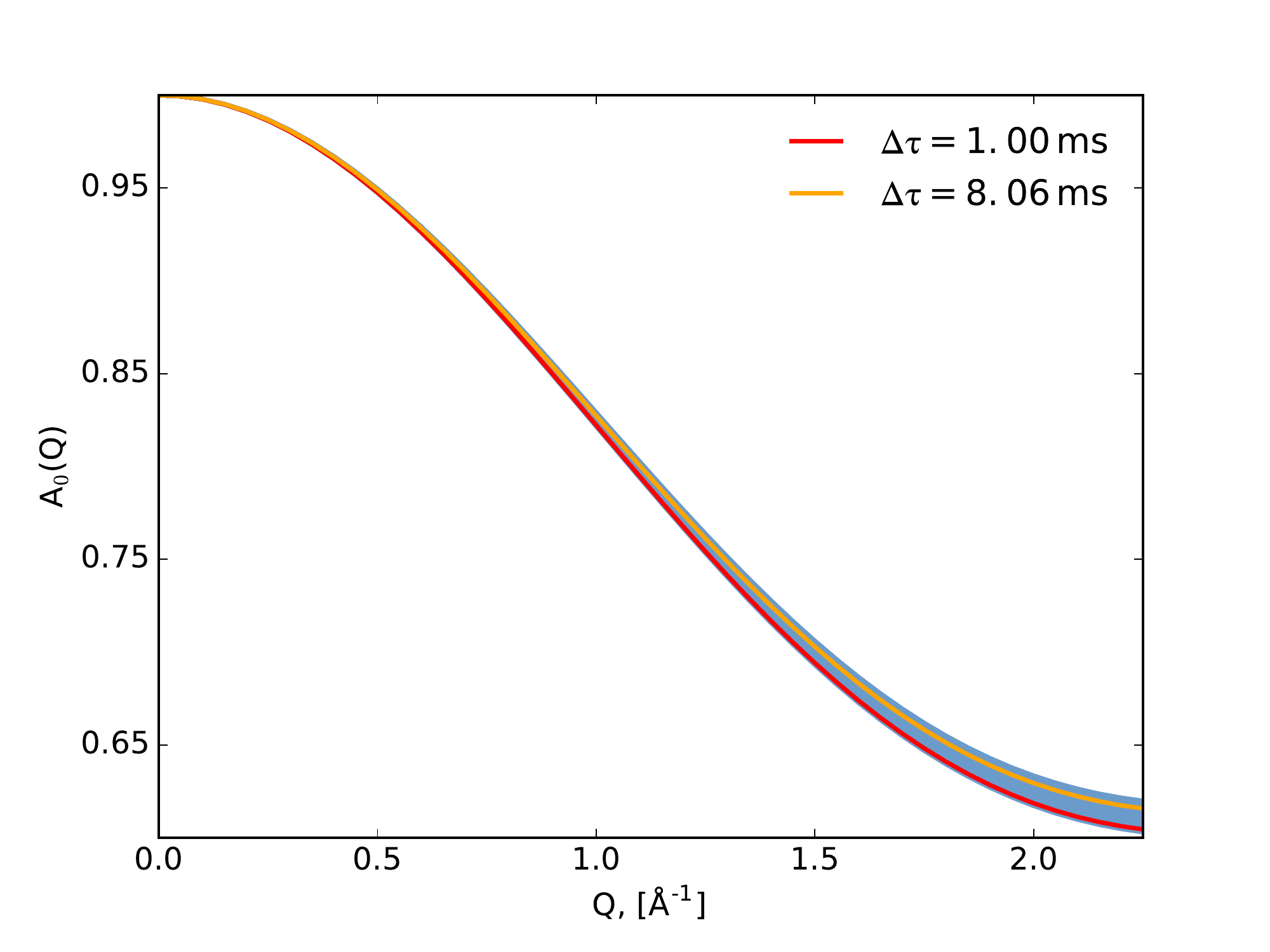}
\caption{EISF as a function of $Q$ at different time delays after laser excitation. $\Delta\tau_1 = 1000\;\mu$s, $\Delta\tau_2 = 8060\;\mu$s are depicted with red and orange color. The blue curves present EISF calculated for $\Delta\tau_i > 75$ ms.}
\label{EISF_Q}
\end{figure}

The analysis of the individual slices showed that some of the adjustable parameters in eq~\ref{SQE2} did not exhibit any statistically significant deviation on the timescale of the photocycle in PM. The relaxation time for rotation ranged from 2.3~ps to 3.5~ps depending on the $d$-spacing (Figure~\ref{tau_d}), but any possible laser induced changes did not exceed 0.4~ps. The parameters describing the energy independent background and the intensity factors remained constant as well. The latter means that the mean square displacement of faster vibrational motion $\langle u^2\rangle$ lies within the error margins of $\pm3\cdot10^{-3}$~\AA$^2$ in the course of the photocycle. Therefore, these parameters were fixed to improve the statistics for the other relevant variables in the further data evaluation. 

The radii and the fraction of mobile particles tended to show some changes during the first milliseconds after the laser excitation, resulting in the drop of $A_0(Q)$ for larger $Q$-values (Figure~\ref{EISF_Q}):

\begin{equation}
A_0(Q)=(1-p_\text{mob})+p_\text{mob}j_0^2(QR)
\label{elastic}
\end{equation}   

\begin{figure}[p]
\centering
\includegraphics[width=0.7\textwidth, clip, trim=0cm 0cm 0.0cm 0.0cm]{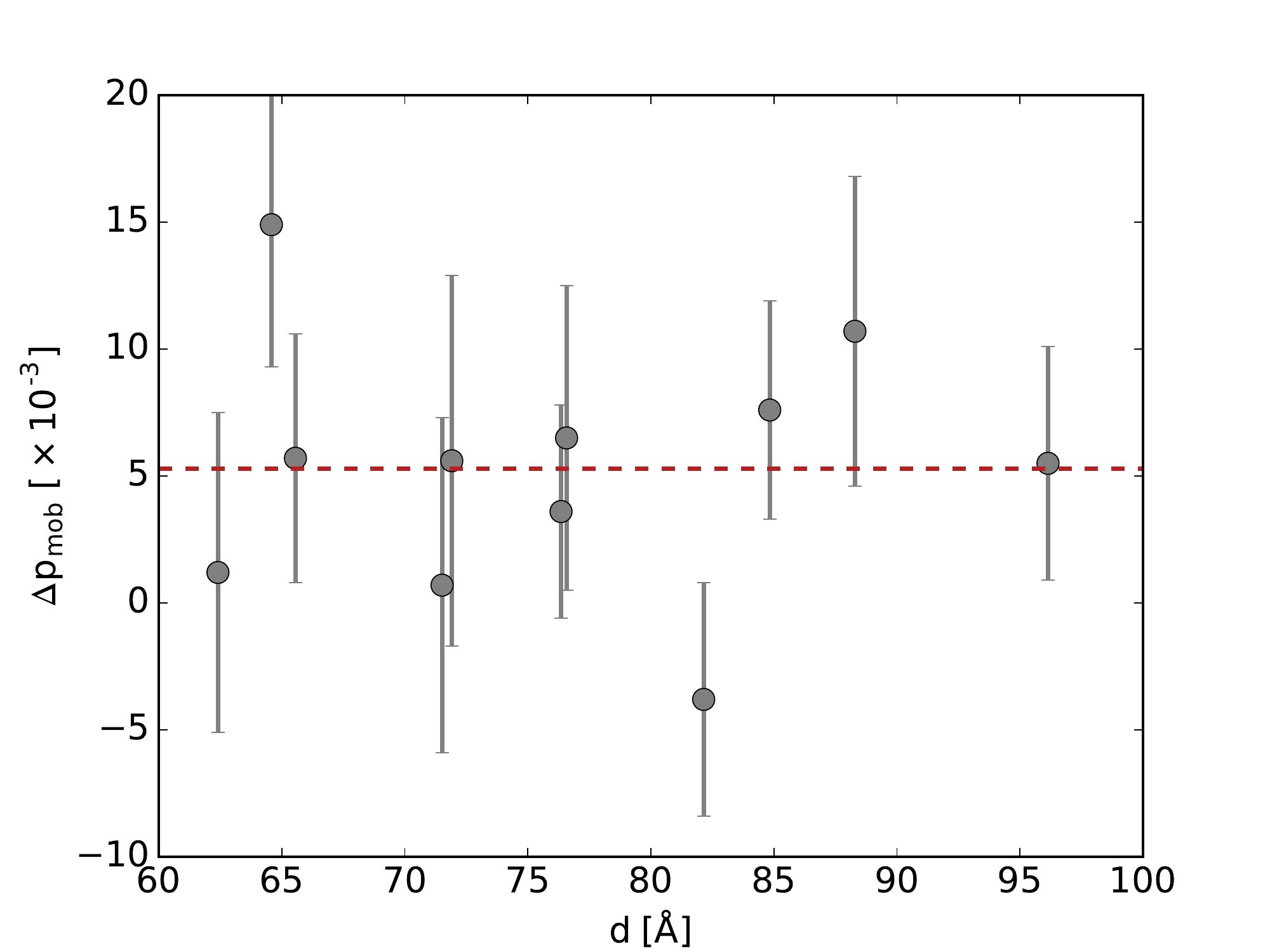}
\caption{Laser induced deviation of the fraction of mobile particles at the time delay of 1000~$\mu$s for the samples with different lamellar lattice constant. The red dashed line corresponds to the value averaged over all samples}
\label{dpmob_d}
\includegraphics[width=0.7\textwidth, clip, trim=0cm 0cm 0.0cm 0.0cm]{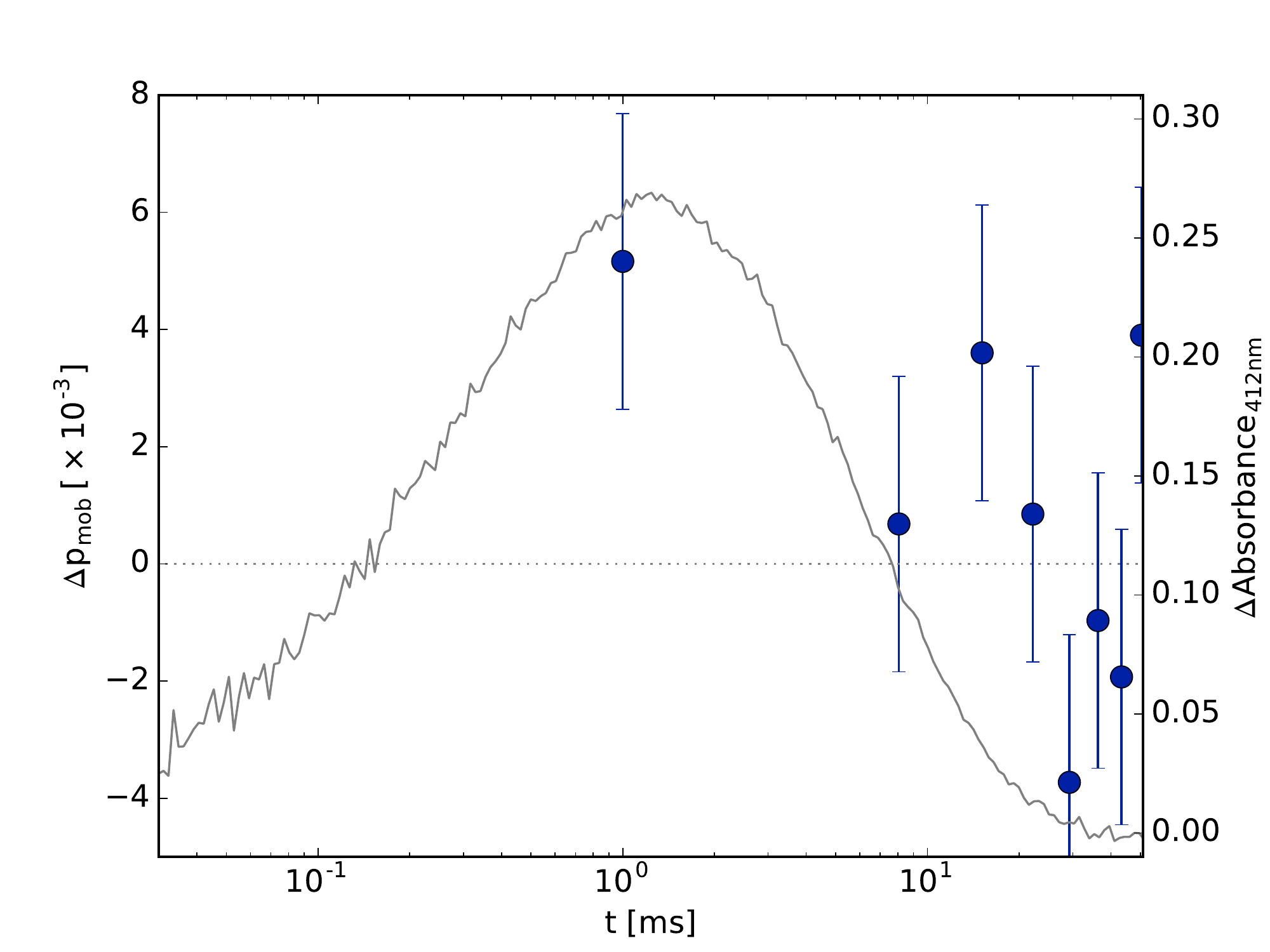}
\caption{Laser induced deviation of the fraction of mobile particles as a function of time delay after the laser excitation. The time delay for the first spectrum equals 1000 $\mu$s. The error bars are 2$\sigma$ of the values calculated for $\Delta t>71$~ms. The gray line is the transient absorption spectrum at $\lambda$=412 nm}
\label{dpmob_50ms}
\end{figure}  

\noindent As can be seen from eq~\ref{elastic}, $p_\text{mob}$ and $R$ are intertwined. In other words, any significant increase in $p_\text{mob}$ may be compensated by a corresponding decrease in $R$. In order to avoid ambiguity in interpreting results, we fixed $R$ to its average value over all slices, especially since the $R$-deviations were both positive and negative with a higher degree of scattering. Finally, on the last step of fitting, only $p_\text{mob}$ remained variable. Additionally, since $p_\text{mob}$ can be considered as an additive parameter, this choice simplifies the analysis of the fit results. In this case it is more justified to compare laser induced deviations $\Delta p_\text{mob}$ evaluated for the samples with different hydration level. Figure~\ref{dpmob_d} depicts this parameter calculated the first spectra of all the samples after the laser excitation ($\Delta \tau$ = 1000 $\mu$s). Unfortunately, due to the large error bars it is impossible to establish any hydration dependence of the laser-induced response of the system, which could be expected, since the photocycle time characteristics are significantly affected by the water content in PM. For example, the maxima of transient absorption changes of the investigated samples are scattered around 1.0$\pm$0.2~ms. Moreover, thermodynamic properties (specific heat capacity, thermal conductivity) of the sample also differ for drier and wetter membrane stacks.  Nevertheless, the average value presented as the red dashed line in Figure~\ref{dpmob_d} corresponds to a statistically significant positive deviation $\Delta p_\text{mob}$, which is in line with the previous work~\cite{Pieper2008, Pieper2009}, where the origin and physiological relevance of the transient modulation of protein dynamics during the M-intermediate of BR have been discussed. According to the interpretation, which also arises from hydration-dependent experiments~\cite{Fitter1999}, the enhanced protein flexibility might function as a "lubricating grease'' for the large-scale structural changes. Moreover, these fast stochastic motions could be a prerequisite to overcome potential barriers of the protein energy landscape during the slower large-scale structural changes~\cite{Frauenfelder1991, HenzlerWildman2007}.  The enhancement of protein thermal fluctuations as a result of light-induced destabilization was also observed by hydrogen/deuterium exchange (HDX) mass spectrometry as well~\cite{Pan2011}. 

Figure~\ref{dpmob_50ms} illustrates how the excess laser-activated fluctuations in the protein-lipid complex change in the course of the time. The deviations of the fraction of mobile particles, $\Delta p_\text{mob}$ are the mean values calculated for all the samples. Although there is a slight enhancement of the picosecond dynamics at $\Delta \tau$=1000~$\mu$s, it is necessary to keep in mind that the changes related to the function of BR during the course of the photocycle can be superimposed with the trivial temperature relaxation of the sample as a result of dissipation of the absorbed light energy. To answer this question unambiguously more time delays up to 5 ms have to be considered, since the temperature relaxation half-decay time is an order of magnitude larger compared to the photocycle length and, thus, the contributions can be discriminated. Moreover, the results of the recent time-resolved wide-angle X-ray scattering study demonstrated significant motions of helices already within 22~$\mu$s of photoactivation, prior to the primary proton transfer event~\cite{Andersson2009}. This highlights the necessity to perform a more careful investigation for other time delays with QENS. However, this is the subject of the upcoming publication.

\section{Conclusion}

The results presented above demonstrate that time-resolved QENS can be a potentially successful technique for studying protein dynamics and unraveling structure-dynamics-function correlations in photosynthesis research.
Performed on samples with different hydration level, the study, however, discloses the existing sensitivity problems of the laser-neutron pump-probe experiment, which requires significantly longer measurements runs to obtain sufficient statistics. Moreover, it has been shown that an exceptional care should be taken to avoid/minimize effects related to the energy dissipation through the sample environment in the time range relevant for the protein function. The next time-resolved QENS study with a more detailed time resolution during the course of the photocycle will demonstrate a more complete picture of the transient modulation of picosecond dynamics in BR. 

\newpage
\bibliography{PMpaper}

\providecommand{\latin}[1]{#1}
\providecommand*\mcitethebibliography{\thebibliography}
\csname @ifundefined\endcsname{endmcitethebibliography}
  {\let\endmcitethebibliography\endthebibliography}{}
\begin{mcitethebibliography}{38}
\providecommand*\natexlab[1]{#1}
\providecommand*\mciteSetBstSublistMode[1]{}
\providecommand*\mciteSetBstMaxWidthForm[2]{}
\providecommand*\mciteBstWouldAddEndPuncttrue
  {\def\EndOfBibitem{\unskip.}}
\providecommand*\mciteBstWouldAddEndPunctfalse
  {\let\EndOfBibitem\relax}
\providecommand*\mciteSetBstMidEndSepPunct[3]{}
\providecommand*\mciteSetBstSublistLabelBeginEnd[3]{}
\providecommand*\EndOfBibitem{}
\mciteSetBstSublistMode{f}
\mciteSetBstMaxWidthForm{subitem}{(\alph{mcitesubitemcount})}
\mciteSetBstSublistLabelBeginEnd
  {\mcitemaxwidthsubitemform\space}
  {\relax}
  {\relax}

\bibitem[Lewandowski \latin{et~al.}(2015)Lewandowski, Halse, Blackledge, and
  Emsley]{Lewandowski2015}
Lewandowski,~J.~R.; Halse,~M.~E.; Blackledge,~M.; Emsley,~L. Direct Observation
  of Hierarchical Protein Dynamics. \emph{Science} \textbf{2015}, \emph{348},
  578--581\relax
\mciteBstWouldAddEndPuncttrue
\mciteSetBstMidEndSepPunct{\mcitedefaultmidpunct}
{\mcitedefaultendpunct}{\mcitedefaultseppunct}\relax
\EndOfBibitem
\bibitem[Henzler-Wildman and Kern(2007)Henzler-Wildman, and
  Kern]{HenzlerWildman2007}
Henzler-Wildman,~K.; Kern,~D. Dynamic Personalities of Proteins. \emph{Nature}
  \textbf{2007}, \emph{450}, 964--972\relax
\mciteBstWouldAddEndPuncttrue
\mciteSetBstMidEndSepPunct{\mcitedefaultmidpunct}
{\mcitedefaultendpunct}{\mcitedefaultseppunct}\relax
\EndOfBibitem
\bibitem[Karplus and McCammon(2002)Karplus, and McCammon]{Karplus2002}
Karplus,~M.; McCammon,~J.~A. Molecular Dynamic Simulations of Biomolecules.
  \emph{Nature Structural \& Molecular Biology} \textbf{2002}, \emph{9},
  646--652\relax
\mciteBstWouldAddEndPuncttrue
\mciteSetBstMidEndSepPunct{\mcitedefaultmidpunct}
{\mcitedefaultendpunct}{\mcitedefaultseppunct}\relax
\EndOfBibitem
\bibitem[Sait\^o(2014)]{Saito2014}
Sait\^o,~H. In \emph{Chapter One - Dynamic Pictures of Proteins by {NMR}};
  Webb,~G.~A., Ed.; Annual Reports on {NMR} Spectroscopy; Academic Press, 2014;
  Vol.~83; pp 1--66\relax
\mciteBstWouldAddEndPuncttrue
\mciteSetBstMidEndSepPunct{\mcitedefaultmidpunct}
{\mcitedefaultendpunct}{\mcitedefaultseppunct}\relax
\EndOfBibitem
\bibitem[Kamerlin and Warshel(2010)Kamerlin, and Warshel]{Kamerlin2010}
Kamerlin,~S. C.~L.; Warshel,~A. At the Dawn of the 21st Century: Is Dynamics
  the Missing Link for Understanding Enzyme Catalysis? \emph{Proteins:
  Structure, Function, and Bioinformatics} \textbf{2010}, \emph{78},
  1339--1375\relax
\mciteBstWouldAddEndPuncttrue
\mciteSetBstMidEndSepPunct{\mcitedefaultmidpunct}
{\mcitedefaultendpunct}{\mcitedefaultseppunct}\relax
\EndOfBibitem
\bibitem[Pieper \latin{et~al.}(2007)Pieper, Hau{\ss}, Buchsteiner,
  Baczy\'{n}ski, Adamiak, Lechner, and Renger]{Pieper2007}
Pieper,~J.; Hau{\ss},~T.; Buchsteiner,~A.; Baczy\'{n}ski,~K.; Adamiak,~K.;
  Lechner,~R.~E.; Renger,~G. Temperature- and Hydration-Dependent Protein
  Dynamics in Photosystem II of Green Plants Studied by Quasielastic Neutron
  Scattering†. \emph{Biochemistry} \textbf{2007}, \emph{46}, 11398--11409\relax
\mciteBstWouldAddEndPuncttrue
\mciteSetBstMidEndSepPunct{\mcitedefaultmidpunct}
{\mcitedefaultendpunct}{\mcitedefaultseppunct}\relax
\EndOfBibitem
\bibitem[Doster and Settles(2005)Doster, and Settles]{Doster2005}
Doster,~W.; Settles,~M. Protein--Water Displacement Distributions.
  \emph{Biochimica et Biophysica Acta (BBA) - Proteins and Proteomics}
  \textbf{2005}, \emph{1749}, 173--186, Solvent Effects\relax
\mciteBstWouldAddEndPuncttrue
\mciteSetBstMidEndSepPunct{\mcitedefaultmidpunct}
{\mcitedefaultendpunct}{\mcitedefaultseppunct}\relax
\EndOfBibitem
\bibitem[Ostermann \latin{et~al.}(2000)Ostermann, Waschipky, Parak, and
  Nienhaus]{Ostermann2000}
Ostermann,~A.; Waschipky,~R.; Parak,~F.~G.; Nienhaus,~G.~U. Ligand Binding and
  Conformational Motions in Myoglobin. \emph{Nature} \textbf{2000}, \emph{404},
  205--208\relax
\mciteBstWouldAddEndPuncttrue
\mciteSetBstMidEndSepPunct{\mcitedefaultmidpunct}
{\mcitedefaultendpunct}{\mcitedefaultseppunct}\relax
\EndOfBibitem
\bibitem[Rasmussen \latin{et~al.}(1992)Rasmussen, Stock, Ringe, and
  Petsko]{Rasmussen1992}
Rasmussen,~B.~F.; Stock,~A.~M.; Ringe,~D.; Petsko,~G.~A. Crystalline
  Ribonuclease A Loses Function below the Dynamical Transition at 220 K.
  \emph{Nature} \textbf{1992}, \emph{357}, 423--424\relax
\mciteBstWouldAddEndPuncttrue
\mciteSetBstMidEndSepPunct{\mcitedefaultmidpunct}
{\mcitedefaultendpunct}{\mcitedefaultseppunct}\relax
\EndOfBibitem
\bibitem[Ernst \latin{et~al.}(2014)Ernst, Lodowski, Elstner, Hegemann, Brown,
  and Kandori]{Ernst2014}
Ernst,~O.~P.; Lodowski,~D.~T.; Elstner,~M.; Hegemann,~P.; Brown,~L.~S.;
  Kandori,~H. Microbial and Animal Rhodopsins: Structures, Functions, and
  Molecular Mechanisms. \emph{Chemical Reviews} \textbf{2014}, \emph{114},
  126--163\relax
\mciteBstWouldAddEndPuncttrue
\mciteSetBstMidEndSepPunct{\mcitedefaultmidpunct}
{\mcitedefaultendpunct}{\mcitedefaultseppunct}\relax
\EndOfBibitem
\bibitem[Haupts \latin{et~al.}(1999)Haupts, Tittor, and Oesterhelt]{Haupts1999}
Haupts,~U.; Tittor,~J.; Oesterhelt,~D. Closing in on Bacteriorhodopsin:
  Progress in Understanding the Molecule. \emph{Annual Review of Biophysics and
  Biomolecular Structure} \textbf{1999}, \emph{28}, 367--399\relax
\mciteBstWouldAddEndPuncttrue
\mciteSetBstMidEndSepPunct{\mcitedefaultmidpunct}
{\mcitedefaultendpunct}{\mcitedefaultseppunct}\relax
\EndOfBibitem
\bibitem[Knoblauch \latin{et~al.}(2014)Knoblauch, Griep, and
  Friedrich]{Knoblauch2014}
Knoblauch,~C.; Griep,~M.; Friedrich,~C. Recent Advances in the Field of
  Bionanotechnology: An Insight into Optoelectric Bacteriorhodopsin, Quantum
  Dots, and Noble Metal Nanoclusters. \emph{Sensors} \textbf{2014}, \emph{14},
  19731--19766\relax
\mciteBstWouldAddEndPuncttrue
\mciteSetBstMidEndSepPunct{\mcitedefaultmidpunct}
{\mcitedefaultendpunct}{\mcitedefaultseppunct}\relax
\EndOfBibitem
\bibitem[Mukhopadhyay \latin{et~al.}(2014)Mukhopadhyay, Cohen, Marchak,
  Friedman, Pecht, Sheves, and Cahen]{Mukhopadhyay2014}
Mukhopadhyay,~S.; Cohen,~S.~R.; Marchak,~D.; Friedman,~N.; Pecht,~I.;
  Sheves,~M.; Cahen,~D. Nanoscale Electron Transport and Photodynamics
  Enhancement in Lipid-Depleted Bacteriorhodopsin Monomers. \emph{ACS Nano}
  \textbf{2014}, \emph{8}, 7714--7722\relax
\mciteBstWouldAddEndPuncttrue
\mciteSetBstMidEndSepPunct{\mcitedefaultmidpunct}
{\mcitedefaultendpunct}{\mcitedefaultseppunct}\relax
\EndOfBibitem
\bibitem[Lanyi(2006)]{Lanyi2006}
Lanyi,~J.~K. Proton Transfers in the Bacteriorhodopsin Photocycle.
  \emph{Biochimica et Biophysica Acta (BBA) - Bioenergetics} \textbf{2006},
  \emph{1757}, 1012--1018, Proton Transfer Reactions in Biological
  Systems\relax
\mciteBstWouldAddEndPuncttrue
\mciteSetBstMidEndSepPunct{\mcitedefaultmidpunct}
{\mcitedefaultendpunct}{\mcitedefaultseppunct}\relax
\EndOfBibitem
\bibitem[Dencher \latin{et~al.}(1989)Dencher, Dresselhaus, Zaccai, and
  B\"uldt]{Dencher1989}
Dencher,~N.~A.; Dresselhaus,~D.; Zaccai,~G.; B\"uldt,~G. Structural Changes in
  Bacteriorhodopsin during Proton Translocation Revealed by Neutron
  Diffraction. \emph{Proceedings of the National Academy of Sciences}
  \textbf{1989}, \emph{86}, 7876--7879\relax
\mciteBstWouldAddEndPuncttrue
\mciteSetBstMidEndSepPunct{\mcitedefaultmidpunct}
{\mcitedefaultendpunct}{\mcitedefaultseppunct}\relax
\EndOfBibitem
\bibitem[Hirai and Subramaniam(2009)Hirai, and Subramaniam]{Hirai2009}
Hirai,~T.; Subramaniam,~S. Protein Conformational Changes in the
  Bacteriorhodopsin Photocycle: Comparison of Findings from Electron and X-Ray
  Crystallographic Analyses. \emph{PLoS ONE} \textbf{2009}, \emph{4},
  e5769\relax
\mciteBstWouldAddEndPuncttrue
\mciteSetBstMidEndSepPunct{\mcitedefaultmidpunct}
{\mcitedefaultendpunct}{\mcitedefaultseppunct}\relax
\EndOfBibitem
\bibitem[Hirai \latin{et~al.}(2009)Hirai, Subramaniam, and Lanyi]{Hirai2009a}
Hirai,~T.; Subramaniam,~S.; Lanyi,~J.~K. Structural Snapshots of Conformational
  Changes in a Seven-Helix Membrane Protein: Lessons from Bacteriorhodopsin.
  \emph{Current Opinion in Structural Biology} \textbf{2009}, \emph{19},
  433--439\relax
\mciteBstWouldAddEndPuncttrue
\mciteSetBstMidEndSepPunct{\mcitedefaultmidpunct}
{\mcitedefaultendpunct}{\mcitedefaultseppunct}\relax
\EndOfBibitem
\bibitem[Andersson \latin{et~al.}(2009)Andersson, Malmerberg, Westenhoff,
  Katona, Cammarata, W{\"o}hri, Johansson, Ewald, Eklund, Wulff, Davidsson, and
  Neutze]{Andersson2009}
Andersson,~M.; Malmerberg,~E.; Westenhoff,~S.; Katona,~G.; Cammarata,~M.;
  W{\"o}hri,~A.~B.; Johansson,~L.~C.; Ewald,~F.; Eklund,~M.; Wulff,~M.
  \latin{et~al.}  Structural Dynamics of Light-Driven Proton Pumps.
  \emph{Structure} \textbf{2009}, \emph{17}, 1265--1275\relax
\mciteBstWouldAddEndPuncttrue
\mciteSetBstMidEndSepPunct{\mcitedefaultmidpunct}
{\mcitedefaultendpunct}{\mcitedefaultseppunct}\relax
\EndOfBibitem
\bibitem[Wickstrand \latin{et~al.}(2015)Wickstrand, Dods, Royant, and
  Neutze]{Wickstrand2015}
Wickstrand,~C.; Dods,~R.; Royant,~A.; Neutze,~R. Bacteriorhodopsin: Would the
  Real Structural Intermediates Please Stand up? \emph{Biochimica et Biophysica
  Acta (BBA) - General Subjects} \textbf{2015}, \emph{1850}, 536--553,
  Structural biochemistry and biophysics of membrane proteins\relax
\mciteBstWouldAddEndPuncttrue
\mciteSetBstMidEndSepPunct{\mcitedefaultmidpunct}
{\mcitedefaultendpunct}{\mcitedefaultseppunct}\relax
\EndOfBibitem
\bibitem[Shibata \latin{et~al.}(2010)Shibata, Yamashita, and Ando]{Shibata2010}
Shibata,~M.; Yamashita,~T. s. K.~H.,~Hayato snd~Uchihashi; Ando,~T. High-Speed
  Atomic Force Microscopy Shows Dynamic Molecular Processes in Photoactivated
  Bacteriorhodopsin. \emph{Nature Nanotechnology} \textbf{2010}, \emph{5},
  208--212\relax
\mciteBstWouldAddEndPuncttrue
\mciteSetBstMidEndSepPunct{\mcitedefaultmidpunct}
{\mcitedefaultendpunct}{\mcitedefaultseppunct}\relax
\EndOfBibitem
\bibitem[B\'ee(1988)]{Bee1988}
B\'ee,~M. \emph{Quasielastic Neutron Scattering, Principles and Applications in
  Solid State Chemistry, Biology and Materials Science}; Adam Hilger, Bristol,
  1988\relax
\mciteBstWouldAddEndPuncttrue
\mciteSetBstMidEndSepPunct{\mcitedefaultmidpunct}
{\mcitedefaultendpunct}{\mcitedefaultseppunct}\relax
\EndOfBibitem
\bibitem[Fitter \latin{et~al.}(2006)Fitter, Gutberlet, and
  Katsaras]{Fitter2006}
Fitter,~J., Gutberlet,~T., Katsaras,~J., Eds. \emph{Neutron Scattering in
  Biology}; Biological and Medical Physics, Biomedical Engineering; Springer,
  2006\relax
\mciteBstWouldAddEndPuncttrue
\mciteSetBstMidEndSepPunct{\mcitedefaultmidpunct}
{\mcitedefaultendpunct}{\mcitedefaultseppunct}\relax
\EndOfBibitem
\bibitem[Gabel \latin{et~al.}(2002)Gabel, Bicout, Lehnert, Tehei, Weik, and
  Zaccai]{Gabel2002}
Gabel,~F.; Bicout,~D.; Lehnert,~U.; Tehei,~M.; Weik,~M.; Zaccai,~G. Protein
  Dynamics Studied by Neutron Scattering. \emph{Quarterly Reviews of
  Biophysics} \textbf{2002}, \emph{35}, 327--367\relax
\mciteBstWouldAddEndPuncttrue
\mciteSetBstMidEndSepPunct{\mcitedefaultmidpunct}
{\mcitedefaultendpunct}{\mcitedefaultseppunct}\relax
\EndOfBibitem
\bibitem[Lechner(2011)]{Lechner2011}
Lechner,~R.~E. \emph{Neutrons in Soft Matter}; John Wiley \& Sons, Inc., 2011;
  pp 203--268\relax
\mciteBstWouldAddEndPuncttrue
\mciteSetBstMidEndSepPunct{\mcitedefaultmidpunct}
{\mcitedefaultendpunct}{\mcitedefaultseppunct}\relax
\EndOfBibitem
\bibitem[Fitter \latin{et~al.}(1999)Fitter, Lechner, and Dencher]{Fitter1999}
Fitter,~J.; Lechner,~R.~E.; Dencher,~N.~A. Interactions of Hydration Water and
  Biological Membranes Studied by Neutron Scattering. \emph{The Journal of
  Physical Chemistry B} \textbf{1999}, \emph{103}, 8036--8050\relax
\mciteBstWouldAddEndPuncttrue
\mciteSetBstMidEndSepPunct{\mcitedefaultmidpunct}
{\mcitedefaultendpunct}{\mcitedefaultseppunct}\relax
\EndOfBibitem
\bibitem[Pieper \latin{et~al.}(2008)Pieper, Buchsteiner, Dencher, Lechner, and
  Hau{\ss}]{Pieper2008}
Pieper,~J.; Buchsteiner,~A.; Dencher,~N.~A.; Lechner,~R.~E.; Hau{\ss},~T.
  Transient Protein Softening during the Working Cycle of a Molecular Machine.
  \emph{Physical Review Letters} \textbf{2008}, \emph{100}, 228103\relax
\mciteBstWouldAddEndPuncttrue
\mciteSetBstMidEndSepPunct{\mcitedefaultmidpunct}
{\mcitedefaultendpunct}{\mcitedefaultseppunct}\relax
\EndOfBibitem
\bibitem[Pieper \latin{et~al.}(2009)Pieper, Buchsteiner, Dencher, Lechner, and
  Hau{\ss}]{Pieper2009}
Pieper,~J.; Buchsteiner,~A.; Dencher,~N.~A.; Lechner,~R.~E.; Hau{\ss},~T.
  Light-induced Modulation of Protein Dynamics During the Photocycle of
  Bacteriorhodopsin. \emph{Photochemistry and Photobiology} \textbf{2009},
  \emph{85}, 590--597\relax
\mciteBstWouldAddEndPuncttrue
\mciteSetBstMidEndSepPunct{\mcitedefaultmidpunct}
{\mcitedefaultendpunct}{\mcitedefaultseppunct}\relax
\EndOfBibitem
\bibitem[Hau{\ss}(2016)]{Hauss2016}
Hau{\ss},~T. Helmholtz-Zentrum Berlin f\"ur Materialien und Energie. (2016).
  V1: Membrane Diffractometer at BER II. \emph{Journal of large-scale research
  facilities} \textbf{2016}, \emph{2}, A94\relax
\mciteBstWouldAddEndPuncttrue
\mciteSetBstMidEndSepPunct{\mcitedefaultmidpunct}
{\mcitedefaultendpunct}{\mcitedefaultseppunct}\relax
\EndOfBibitem
\bibitem[Lechner \latin{et~al.}(1998)Lechner, Fitter, Dencher, and
  Hau{\ss}]{Lechner1998}
Lechner,~R.~E.; Fitter,~J.; Dencher,~N.~A.; Hau{\ss},~T. Dehydration of
  biological membranes by cooling: an investigation on the purple
  membrane11Edited by J. Karn. \emph{Journal of Molecular Biology}
  \textbf{1998}, \emph{277}, 593--603\relax
\mciteBstWouldAddEndPuncttrue
\mciteSetBstMidEndSepPunct{\mcitedefaultmidpunct}
{\mcitedefaultendpunct}{\mcitedefaultseppunct}\relax
\EndOfBibitem
\bibitem[Richard \latin{et~al.}(2013)Richard, Ferrand, Kearley, and
  Bradley]{lamp}
Richard,~D.; Ferrand,~M.; Kearley,~G.~J.; Bradley,~A.~D. The {L}amp Book. 2013;
  \url{http://www.ill.eu/?id=2024}\relax
\mciteBstWouldAddEndPuncttrue
\mciteSetBstMidEndSepPunct{\mcitedefaultmidpunct}
{\mcitedefaultendpunct}{\mcitedefaultseppunct}\relax
\EndOfBibitem
\bibitem[Burankova(2014)]{BurankovaPhD}
Burankova,~T. Dynamics and Structure of Ionic Liquids by Means of Neutron
  Scattering. Ph.D.\ thesis, Saarland University, 2014\relax
\mciteBstWouldAddEndPuncttrue
\mciteSetBstMidEndSepPunct{\mcitedefaultmidpunct}
{\mcitedefaultendpunct}{\mcitedefaultseppunct}\relax
\EndOfBibitem
\bibitem[Markwardt(2009)]{Markwardt2009}
Markwardt,~C.~B. Non-Linear Least Squares Fitting in {IDL} with {MPFIT}.
  \emph{Astronomical Data Analysis Software and Systems {XVIII}} \textbf{2009},
  \emph{411}, 251--254\relax
\mciteBstWouldAddEndPuncttrue
\mciteSetBstMidEndSepPunct{\mcitedefaultmidpunct}
{\mcitedefaultendpunct}{\mcitedefaultseppunct}\relax
\EndOfBibitem
\bibitem[Tobias \latin{et~al.}(2009)Tobias, Sengupta, and Tarek]{Tobias2009}
Tobias,~D.~J.; Sengupta,~N.; Tarek,~M. Hydration Dynamics of Purple Membranes.
  \emph{Faraday Discussions} \textbf{2009}, \emph{141}, 99--116\relax
\mciteBstWouldAddEndPuncttrue
\mciteSetBstMidEndSepPunct{\mcitedefaultmidpunct}
{\mcitedefaultendpunct}{\mcitedefaultseppunct}\relax
\EndOfBibitem
\bibitem[Hempelmann(2000)]{Hempelmann2000}
Hempelmann,~R. \emph{Quasielastic Neutron Scattering and Solid State
  Diffusion}; Oxford series on neutron scattering in condensed matter;
  Clarendon {P}ress, {O}xford, 2000\relax
\mciteBstWouldAddEndPuncttrue
\mciteSetBstMidEndSepPunct{\mcitedefaultmidpunct}
{\mcitedefaultendpunct}{\mcitedefaultseppunct}\relax
\EndOfBibitem
\bibitem[Volino \latin{et~al.}(2006)Volino, Perrin, and Lyonnard]{Volino2006}
Volino,~F.; Perrin,~J.-C.; Lyonnard,~S. Gaussian Model for Localized
  Translational Motion: Application to Incoherent Neutron Scattering. \emph{The
  Journal of Physical Chemistry B} \textbf{2006}, \emph{110},
  11217--11223\relax
\mciteBstWouldAddEndPuncttrue
\mciteSetBstMidEndSepPunct{\mcitedefaultmidpunct}
{\mcitedefaultendpunct}{\mcitedefaultseppunct}\relax
\EndOfBibitem
\bibitem[Frauenfelder \latin{et~al.}(1991)Frauenfelder, Sligar, and
  Wolynes]{Frauenfelder1991}
Frauenfelder,~H.; Sligar,~S.~G.; Wolynes,~P.~G. The Energy Landscapes and
  Motions of Proteins. \emph{Science} \textbf{1991}, \emph{254},
  1598--1603\relax
\mciteBstWouldAddEndPuncttrue
\mciteSetBstMidEndSepPunct{\mcitedefaultmidpunct}
{\mcitedefaultendpunct}{\mcitedefaultseppunct}\relax
\EndOfBibitem
\bibitem[Pan \latin{et~al.}(2011)Pan, Brown, and Konermann]{Pan2011}
Pan,~Y.; Brown,~L.; Konermann,~L. Hydrogen Exchange Mass Spectrometry of
  Bacteriorhodopsin Reveals Light-Induced Changes in the Structural Dynamics of
  a Biomolecular Machine. \emph{Journal of the American Chemical Society}
  \textbf{2011}, \emph{133}, 20237--20244\relax
\mciteBstWouldAddEndPuncttrue
\mciteSetBstMidEndSepPunct{\mcitedefaultmidpunct}
{\mcitedefaultendpunct}{\mcitedefaultseppunct}\relax
\EndOfBibitem
\end{mcitethebibliography}

\end{document}